**Ethyl cellulose-based thermoreversible organogel photoresist for sedimentation-free volumetric additive manufacturing**

*Joseph T. Toombs, Ingrid K. Shan, Hayden K. Taylor\**

J.T.T, I.K.S, H.K.T.
Department of Mechanical Engineering, University of California, Berkeley, CA 94720, United States of America.
E-mail: hkt@berkeley.edu

Liquid photoresists are abundant in the field of light-based additive manufacturing (AM). However, printing unsupported directly into a vat of material in emerging volumetric AM technologies—typically a benefit due to fewer geometric constraints and less material waste—can be a limitation when printing low-viscosity liquid monomers and multimaterial constructs due to part drift or sedimentation. With ethyl cellulose (EC), a thermoplastic soluble in organic liquids, we formulate a simple three-component transparent thermoreversible gel photoresist with melting temperature of ~64 °C. The physically crosslinked network of the gel leads to storage moduli in the range of 0.1–10 kPa and maximum yield stress of 2.7 kPa for a 10 wt% EC gel photoresist. Non-zero yield stress enables sedimentation-free tomographic volumetric patterning in low-viscosity monomer without additional hardware or modification of apparatus. Additionally, objects inserted into the print container can be suspended in the gel material which enables overprinting of multimaterial devices without anchors connecting the object to the printing container. Flexural strength is also improved by 100% compared to the neat monomer for a formulation with 7 wt% EC.

# 1. Introduction

Photoresists for light-based additive manufacturing (AM) have traditionally had liquid state of matter. For stereolithography, a layer-by-layer light-based AM process,[1,2] the liquid state is advantageous because it allows fresh material to flow into the space between the base of the tray containing it and the emerging partially fabricated part. Only after this replenishment or reflow can printing of the next layer commence.

Volumetric additive manufacturing (VAM) techniques like direct laser writing (DLW) by two photon polymerization[3] (2PP), light-sheet-based printing,[4,5] and computed axial lithography, also known as tomographic VAM[6–9], have made it possible to print directly within a volume of precursor material.[10] Volumetric patterning avoids layer replenishment; however, liquid photoresists are still commonly used for VAM techniques because they are abundant in the light-based AM field. In low-viscosity photoresists, the topology of structures written by DLW can be restricted because noncontiguous features may drift in the liquid during stage scanning or stitching moves and unsupported features may sag.[11,12] In tomographic VAM, the crosslinked network densifies upon gelation and sedimentation may therefore occur before exposure is complete since structures are often printed without anchors to a substrate.

Dry-film photoresists (solid-state photoresists) have a long history in the field of semiconductor lithography due to the decreased diffusion of reactive species which increases material contrast and improves minimum feature size.[13] Typical photoresists consist of high-concentration solid polymer and solvent (to dissolve the polymer and produce a liquid that may be casted onto the substrate). The solvent is evaporated out of the liquid after deposition, resulting in a solid layer into which 3D features can be written with a standard photolithography aligner, holographically, or with DLW.[14,15] Sol-gel composites which gel after thermal treatment and evaporation of solvents used in synthesis have also been developed for DLW.[16] These approaches are not readily applicable to centimeter-scale AM because solvent evaporation quickly becomes impractical for large film thicknesses and bulk volumes.[13]

An alternative approach is to employ a reversibly crosslinking polymer. The Diels-Alder reaction between furan and maleimide moieties forms a thermally reversible covalent crosslink. When combined with a thiol-ene or acrylate crosslinking system, the Diels-Alder dual-cure material enables printing of challenging unsupported microscale features with DLW.[17,18] However, the temperature-dependent Diels-Alder equilibrium requires elevated temperatures to maintain conversion above the gelation point (to remain in gel state) and thus also special modifications to the light exposure apparatus.

Whereas the Diels-Alder reaction is a reversible chemical crosslink, reversible physical crosslinks, i.e., noncovalent interactions, offer an alternative mechanism to attain a gel-state photoresist. Thermally reversible gelatin-based hydrogels have been utilized in VAM, including for production of biological constructs and shape-memory materials.[7,19–22] Particularly for bioprinting, light delivery occurs while the precursor material is in gel state which prevents sedimentation of embedded cells and after printing supports the polymerized structure. However, by design, these aqueous materials are soft after polymerization and do not possess the mechanical properties required by many engineering applications. On the other hand, thermoplastic binders added to monomer mixtures have been utilized to record volume holograms as the spatial variation in density of sensitizers or in refractive index of the affected material for data storage applications, albeit, typically in thin films.[23]

In each case, whether chemically or physically crosslinked, photopolymerization of the primary photoactive network "fixes" reversible crosslinks of the secondary matrix. Post-exposure, the material is heated (or cooled depending on the thermal dependence of the solid-to-liquid phase transition[22]) to destroy the secondary network in the unfixed, or uncured, regions and the printed structure is extracted.

In this work, we introduce a thermally reversible gel-state organic photoresist based on cellulose-derived ethyl cellulose. The simple formulation that we propose enables fast solventless preparation of a transparent, room-temperature, nonaqueous gel photoresist which makes VAM of challenging geometries and multi-material constructs possible with low-viscosity organic monomers.

## 2. Results/Discussion

Ethyl cellulose (EC) is a polymer derived by etherification of cellulose in which some or all of the hydroxyl groups of cellulose are substituted with ethyl groups.[24] EC may be produced with variable degree of substitution (DS) of the hydroxl groups on the anhydroglucose unit (maximum of 3) which affects the solubility in oils and aqueous solutions. At low DS (<2.0) it is soluble in water while at high DS (2.6–2.8) it is soluble in organic solvents.[25] The ability to dissolve in oils has utility in food and nutrition research to produce structured oils, oleogels, which mimic the texture of natural fats and could be a substitute for saturated fats in food products.[26,27] In the pharmaceutical industry, EC has been used extensively as an excipient for oral and transdermal drug delivery.[28]

EC forms a physical gel when it is dissolved in certain organic liquids, e.g., aromatic hydrocarbons, esters, and acetates,[29] and unsubstituted hydroxyl groups interact through dipole-dipole interactions with polar sites on organic solvents[30,31] (see supplementary materials table S1 which gives solubility ≤10 wt% in several common monomers). In this work, we used trimethylolpropane triacrylate (TMPTA) as a homopolymerizing "solvent" in which EC was dissolved. The viscosity of TMPTA is ~0.1 Pa s. Preparation of the photoresist gel consisted of three steps (**Figure 1**): 1) EC was mixed with the TMPTA monomer at an elevated temperature of 125 °C, near the glass transition temperature of EC,[32] until complete dissolution. 2) The temperature of the solution was reduced to 80 °C and photoinitiator (PI) was added (see Experimental Section). 3) The solution was dispensed at 80 °C into containers for tomographic VAM and allowed to cool to room temperature forming a gel which was stable for ≥3 weeks and free-standing in films ≥5 mm in thickness and bulk bodies ≥40 mm in height. In the experiments that follow, four formulations with varying EC concentration (in wt%) were typically investigated: 0 (TMPTA only), 4, 7, and 10. 4 wt % EC was selected as the minimum non-zero concentration because it was the minimum required to create a gel which passed the tube inversion test.

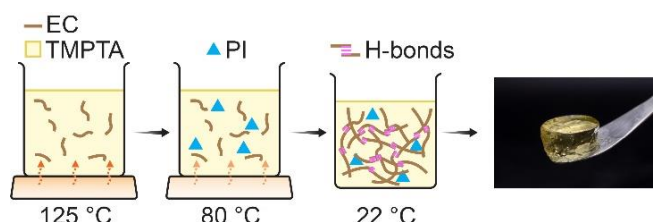

**Figure 1.** Preparation of the ethyl cellulose-based organogel photoresist.

Unlike vat polymerization or stereolithography, which absorbs light in a thin layer at an interface, VAM requires that light can penetrate the volume of photoresist without significant absorption. Therefore, the gel-state photoresist must have sufficient transmission at the writing wavelength to support selective polymerization within the volume. We evaluated transmission of the gel with UV-vis spectroscopy at wavelengths of 300–750 nm (**Figure 2**). Transmission was greater than 50% for a 1 cm pathlength of 10 wt% EC photoresist for wavelengths greater than 405 nm. In printing experiments shown later, 442 nm and 455 nm light sources were used to maximize transmission while still overlapping the photoinitiator absorption peak. Acceptable transmission ≥405 nm suggests that the gel photoresist can be used for techniques utilizing either single photon polymerization in the near ultraviolet range e.g., macro and micro-VAM and light-sheet printing, or 2PP in the near infrared range e.g., DLW.

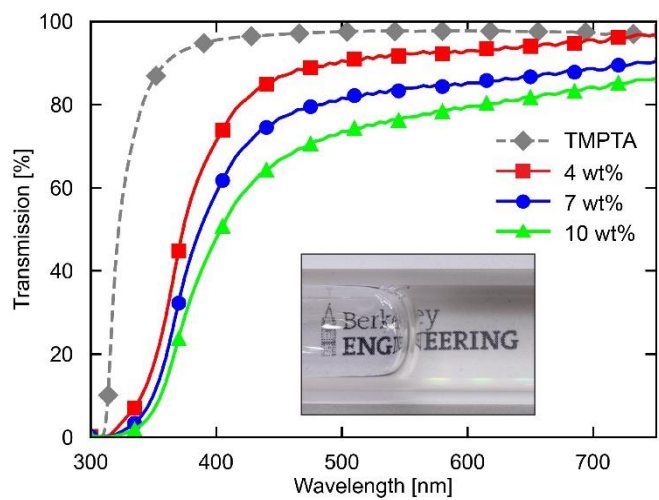

**Figure 2.** EC gel photoresist UV-vis spectroscopy. Transmission % as a function of EC concentration in wt%. The inset shows a 1 cm cuvette filled with 10 wt% EC in TMPTA over text.

The melt characteristics of the gel photoresist are important for identification of liquid processing temperatures needed during gel preparation and post-polymerization development. The melt characteristics were investigated with thermorheology studies. In **Figure 3a**, the storage ($G'$) and loss modulus ($G''$) are plotted as a function of temperature. Moduli were measured during application of increasing temperature at a constant rate of 0.17 °C s$^{-1}$. At ambient temperature, the elastic properties of the gel are correlated with the concentration of EC with 10 wt% having the largest storage modulus and smallest loss tangent ($\tan \delta = G''/G'$) and 4 wt% the smallest storage modulus and largest loss tangent. The melting temperature was estimated as the temperature at which the loss tangent becomes greater than 1 or the crossover point of G′ and G″. The melting temperature ~64 °C does not change substantially with variation in EC concentration. This agrees with previous studies of EC gels, which concluded that the melting or softening point is more strongly dependent on the molecular weight and not significantly dependent on the concentration of EC.[33,34]

The monotonically decreasing modulus with increasing temperature (along with variable concentration) allows for careful tuning of the viscoelastic properties for extrusion and deposition applications. This property—which is likely due to intermolecular hydrogen bonding between EC chains without secondary conformations (for DS ~2.5)—is uncommon among thermoreversible gels.[33,35]

Others such as gellan gum, agarose, and kappa carrageenan hydrogels exhibit step-like melting behavior originating from helical aggregation of polymer chains.[33] Although the primary function of the photoresist gel explored in this work is to support sedimentation-free VAM, the thermo-rheological properties of the gel suggest utility in extrusion-based or direct ink writing AM techniques.

At room temperature the material has gel state, but the liquid state of entrapped monomer permits photopolymerization. We investigated the effects EC had on kinetics of photopolymerization in terms of shear modulus with photorheology (Figure 3b). The onset of polymerization was not affected substantially comparing between 0 wt% and non-zero concentrations as well as between different EC concentrations. The rate of polymerization was reduced with increased EC concentration which could be due to relatively slower propagation rates in more dilute monomer concentrations (a result of EC present in the mixture). However, the ultimate shear modulus was not affected substantially.

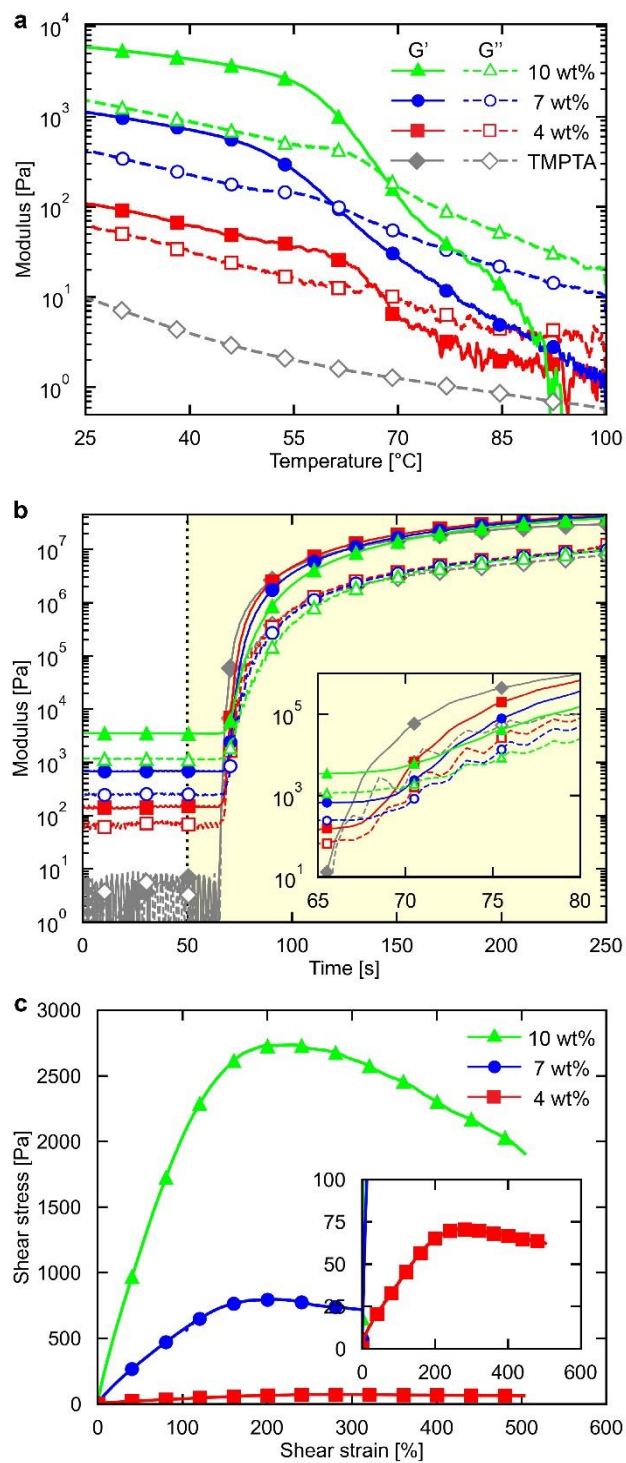

**Figure 3.** Gel photoresist rheology: Storage and loss modulus for four formulations of 0 (TMPTA only), 4, 7, and 10 wt% EC, (a) as a function of temperature and (b) as a function of time under illumination (light is activated after the dashed vertical line in the shaded region). (c) Shear stress vs shear strain for 4, 7, and 10 wt% EC. The peak stress of each curve represents the yield stress of the gel.

The yield stress of the gel was measured with a shear stress ramp and identification of the peak stress (Figure 3c). The shear rate was set to 0.01 s$^{-1}$ to simulate near static loading conditions which are most similar to sedimentation of a printed object. Increase in yield stress was correlated with EC concentration and the 10 wt% EC gel exhibited the largest yield stress of 2.7 kPa. We expect this trend to continue as EC concentration is increased, however, >10 wt% was difficult to dissolve in the monomer used here and the yield stress ≤10 wt% was sufficient to prevent sedimentation of printed objects at cm scale or smaller.

The utility of non-zero yield stress is in preservation of print fidelity with low viscosity precursors. In general, patterned light exposure must continue for a short period after the polymer gelation point to reach a green state which is structurally stable to resist deformation and avoid collapse during development. However, in a low-viscosity monomer, the partially formed object begins to sink due to densification after gelation and the latent image of the 3D light dose is distorted as the object (and a boundary layer of preexposed monomer) moves relative to the spatially static digital light projections (**Figure 4 a–c**). Latent image VAM, in which the object is instead flood-exposed during the phase of photopolymerization between gelation and sufficient polymerization has been proposed as a solution to this sedimentation problem.[36] However, this technique requires: 1) a strongly nonlinear relationship—which is sometimes unattainable—between the dose and the polymer degree of conversion to prevent unwanted polymerization in the background; 2) a high-power diffuse light source to illuminate uniformly the entire print container. Because the photoresist gel in this work had non-zero yield stress, it supported the partially polymerized object during the post-gelation illumination phase. The patterned exposure phase could be completed without relative motion between the light patterns and the printed object, and sufficient polymer conversion was reached without the distortion that was observed in the 0 wt% EC experiment (Figure 4 d–f) and without an additional light source.

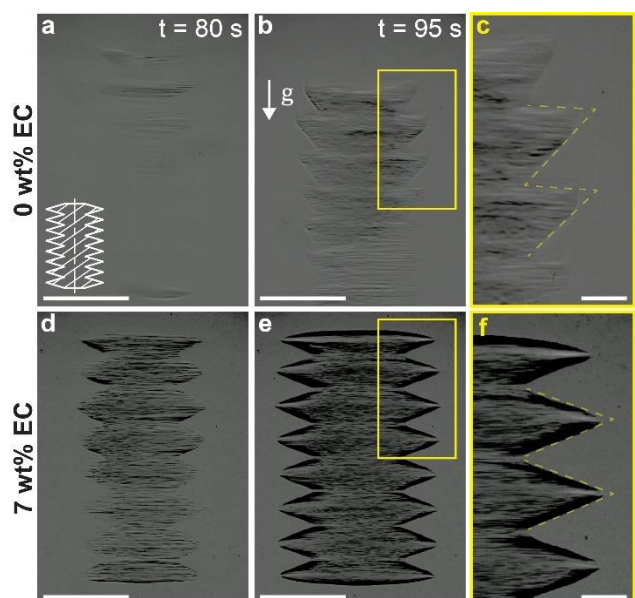

**Figure 4.** In-situ imagery showing sedimentation prevention with gel photoresist. An axisymmetric structure with isosceles triangular features (design cross section pictured in inset of a) was printed with tomographic VAM in (a, b) TMPTA (0 wt% EC) and (d, e) 7 wt% EC gel. $\vec{g}$ indicates the direction of the gravitational vector. Scale bars: 1 mm. (c, f) Insets show magnified surface structure. Without EC, as the object sinks, the voids between protrusions are overexposed and the protrusions become sawtooth shaped. With EC, the object does not sink, and isosceles triangular features are preserved. Scale bars: 0.25 mm.

Overprinting is a fabrication method enabled by tomographic VAM in which an object is 3D printed onto a preexisting object or *insert* in the print volume. Not only can the gel state of the EC photoresist prevent printed objects from sinking but it also allows objects to be suspended in the volume without support or anchors to the print container or external apparatus. We fabricated a mock auditory device to demonstrate how this feature of the gel is advantageous for overprinting. The overprinted geometry which encapsulated an induction LED and "circuit board" was designed to custom-fit a model of a human ear (**Figure 5a**).

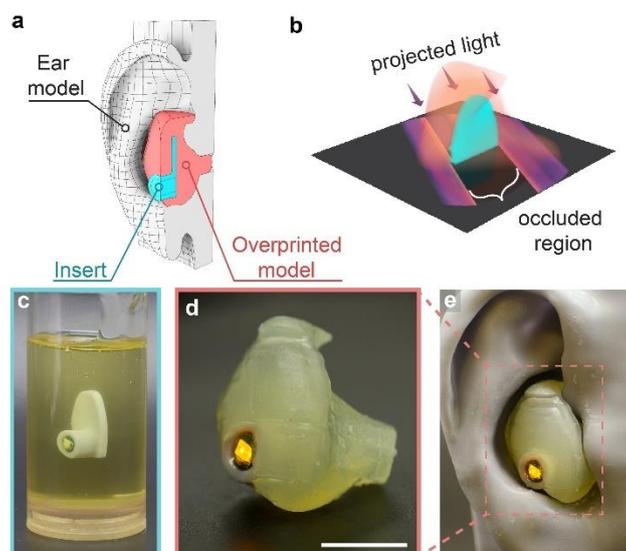

**Figure 5.** Overprinting with tomographic VAM and EC gel photoresist. (a) CAD model of overprinted auditory device. (b) Projected light is occluded by the insert which is accounted for in forward and backprojection computations. (c) Insert suspended in gel photoresist prior to printing. (d) Overprinted model. Scale bar: 10 mm. (e) Auditory device fitted into ear canal and outer ear of 3D printed model.

Digital light projections were computed and optimized with the object-space model optimization algorithm[37] using forward and inverse models which account for light occlusion by the insert[38] (Figure 5b). The insert was suspended in the gel (Figure 5c) in a two-step process: first, a small amount of liquified 7 wt% gel was dispensed into the container and allowed to cool; second, the insert was embedded in the gel and more liquified gel was dispensed to cover the insert completely. After printing, the gel was heated past its melting point, excess liquid was poured off, and followed by solvent development in isopropyl alcohol. After completion, the auditory device (Figure 5d) was fitted into the ear model with good shape conformation (Figure 5e).

Finally, we investigated the effect EC has on the mechanical strength of casted flexural specimen following the ASTM D790 specifications (**Figure 6**). Without EC additive, the polymerized material had ultimate flexural strength of 12.8 MPa. Increased EC content increased ultimate flexural strength up to a maximum of 25.6 MPa with 7 wt%, an increase of 100 % over 0 wt%. Above 7 wt%, the mean flexural strength decreased, which suggests there may be an optimal EC concentration (to improve flexural strength) in the range of 4–10 wt%. However, the observed decrease in flexural strength from 7 to 10 wt% was not statistically significant at the 5% level; therefore, further testing is required to confirm this hypothesis. Overall, the increased strength could be attributed to residual EC in the crosslinked polymer network acting as a reinforcing filler in the TMPTA-EC composite. Indeed, cellulosic materials have been widely used as biobased fillers in thermoset composites to modify mechanical properties.[39] For instance, the increase in strength we observed is consistent with literature where EC was used as a filler in crystalline polymers.[40] Without chemical modification of the monomer or filler, the effect of filler average molecular weight on the strength of the composite could be studied. On the other hand, grafting photocrosslinkable moieties (e.g., acrylics and alkenes/ynes (with thiol crosslinker)) onto the cellulose polymer chain to produce a thermoreversible gelling agent which is active in the photopolymerization reaction may further improve mechanical strength.[41]

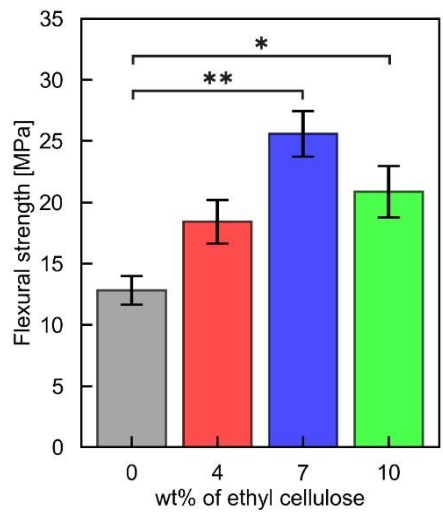

**Figure 6.** Mean flexural strength of polymerized gel. n = 8 samples were tested for each concentration of EC. Error bars represent mean ± one standard deviation. p-values were calculated with Welch's t-test. * $p < 0.05$, ** $p < 0.01$.

## 3. Conclusion

Ethyl cellulose can be used to create a gel-state photoresist without chemical modification of the primary monomer(s), which results in simple formulations and preparation. Thermoreversibility at low temperatures provides an accessible mechanism to process a gel-state material after VAM patterning. This approach is not limited to homopolymerizing material systems as used in this work. It could be used with heteropolymer and hybrid material systems such as epoxy–vinyl systems[42], which may otherwise be unprintable with VAM due to insufficient viscosity, as long as EC is soluble in the constituents. Investigation of how solubility in monomers is affected by varying ethyl DS or grafting of photocrosslinkable functionalities on cellulose or perhaps alternative thermoreversible gelators (e.g., cellulose acetate, or non-cellulose-based polymers) will be important to expand the list of compatible monomer–gelator systems. In the work presented here, the mechanical properties improved as higher flexural strength was achieved with an EC–thermoset composite compared to only thermoset; however, this may not always be the case, especially if higher EC concentration is required to create a gel in very low-viscosity monomers. In future research, it will be important to investigate methods to achieve gel characteristics with lower gelator concentration (e.g., tuning the molecular weight of EC) to minimize influence on optical and mechanical properties and photopolymerization kinetics of the neat material system. Overall, we demonstrated the utility of a gel-state organic photoresist for tomographic VAM printing of low-viscosity monomer without sedimentation and for overprinting on freely suspended objects for individualized and functionally integrated devices. We expect this methodology will enable new applications for light-sheet based printing and tomographic VAM such as single-step fabrication of soft robotics with embedded pneumatics and electronics.

**4. Experimental Section/Methods**

*Materials:* Trimethylolpropane triacrylate was purchased from TCI America. Ethyl cellulose (48–49% ethoxy basis, 30–70 mPa·s 5% in toulene), camphorquinone (CQ) photoinitiator, ethyl 4-(dimethylamino)benzoate (EDAB), and 2,2,6,6-Tetramethylpiperidine 1-oxyl (TEMPO) radical scavenger was purchased from Sigma Aldrich. All other monomers besides TMPTA used to evaluate solubility were purchased from Sigma Aldrich. CQ was added at 5 mM concentration for all photopolymerization experiments. CQ and co-initiator EDAB were used together in 1:1 wt CQ/wt EDAB combination for all photopolymerization experiments.

*UV-VIS analysis:* Heated liquid samples of photoresist with varying concentrations of EC were dispensed into 1 cm path-length quartz cuvettes. The cuvette and EC photoresist was allowed to cool to room temperature by natural (unforced) cooling over a 24-hour period. Samples were measured with reference to air in an empty quartz cuvette on a Shimadzu UV-2600 Plus spectrophotometer. The recorded transmission was corrected by accounting for Fresnel reflection at the air–quartz interface in the reference arm of the measurement. A dispersion relation[43] was used for the refractive index of the cuvette.

*Thermorheology analysis:* All rheological studies were performed on an Anton Paar Physica MCR301 rheometer with 8 mm diameter upper parallel plate and heated lower plate. Oscillatory shear loading was performed on each gel and the neat monomer. Gel samples had 8 mm diameter and 1.8 mm thickness. The parallel plate gap was set to 0.3 mm when testing the neat monomer. For all thermorheology experiments, the shear strain oscillation was constant with an amplitude 1% strain and frequency of 5 Hz. Temperature was increased from 25 °C to 100 °C at a constant rate of 0.17 °C s$^{-1}$.

*Photorheology analysis:* Oscillatory shear loading was performed on each gel formulation and the neat monomer. Gel samples had 8 mm diameter and 1 mm thickness. The parallel plate gap was set to 0.3 mm when testing the neat monomer. Oscillation amplitude and frequency settings used in thermorheology experiments were also used here. A custom LED attachment (center wavelength of 443 nm) was affixed to the lower plate of the rheometer. An LED was located underneath a microscope slide on top of which the gel or liquid to be analyzed was dispensed. Each measurement was performed on a fresh microscope slide. The custom attachment allowed light to pass through the glass slide and polymerize the material while preventing motion of the slide and allowing oscillatory measurement of the storage and loss moduli over time. Before switching on the LED, moduli were measured for a 50 s period under no illumination.

*Yield stress rheological analysis:* Shear stress ramps were performed on each gel formulation with samples of 0.7 mm thickness and 8 mm diameter. The constant shear strain rate was set to 0.01 s$^{-1}$. The upper and lower plates were covered with a large grit sandpaper to prevent wall slippage at high shear strain.

*Mechanical strength analysis:* Testing coupons were created using a custom polydimethylsiloxane mold following the ASTM D790 specifications. The EC gel photoresist was heated to a liquid state then cast in the mold and left in a dark area for 24 hours to allow it to cool to room temperature and return to gel state. Then, the photoresist was cured in a FormLabs Form Cure for 40 mins to ensure complete polymerization.

Three-point flexural tests were performed on a MARK-10 ESM1500 with a Model 5i force gauge and MR01-200 force sensor (maximum load 1 kN). Coupons were strained at 0.1 mm mm$^{-1}$ min$^{-1}$ per Procedure B of ASTM D790 Section 4.

*Tomographic VAM printing:* For tomographic VAM printing, 1.5 mM TEMPO was added to the formulation to improve printing contrast. Heated liquid EC photoresist was dispensed into 40 mL vials and 1 mL vials for cm-scale (Figure 5) and μm-scale tomographic VAM (Figure 4), respectively. For overprinted parts, the insert on which the part was printed was placed into the vial in this stage. The vials were then set in a dark area for 24 hours to cool to room temperature prior to printing. The desired geometry was then printed with the gel-state EC photoresist with no modifications to the printing procedure.

After the desired geometry was printed, vials were heated in a water bath to liquefy the surrounding unpolymerized gel. Excess liquid was then poured out and the printed part was rinsed in isopropyl alcohol for three minutes. The part was then post-cured for five minutes in a FormLabs Form Cure chamber.

Computed axial lithography apparatuses described in previous works were used for cm-scale[44] and μm-scale[9] printing. In the apparatus for cm-scale printing, the light source was changed to a 455 nm LED.

*Statistical analysis:* In rheology analyses, n = 1 samples were tested for each condition. Data presented is actual recorded data.

In mechanical strength analysis, for each sample type, n = 8 samples were tested. Outliers were removed with the interquartile range method on the peak flexural stress at failure. Data in Figure 6 is presented as mean ± one standard deviation. A two-sided Welch's t-test was used to assess significance of differences in the mean ultimate flexural stress between sample types. Alpha significance levels of 0.05 and 0.01 were used to classify statistical significance and strong statistical significance, respectively. Microsoft Excel was used for this statistical analysis.

**Acknowledgements**

The authors would like to thank the lab of Professor Sanjay Kumar lab in the Department of Bioengineering at UC Berkeley for use of the rheometer. This work was funded by the National Science Foundation under cooperative agreement no. EEC-1160494.

# Supporting Information

**Ethyl cellulose-based thermoreversible organogel photoresist for sedimentation-free volumetric additive manufacturing**

Joseph T. Toombs, Ingrid K. Shan, Hayden K. Taylor*

Table S1. Solubility of EC in various monomers

| Monomer | Abbreviation | Solubility in wt % | Chemical structure of monomer |
|---|---|---|---|
| Neopentyl glycol dimethacrylate | NGDMA | ≥ 10 | 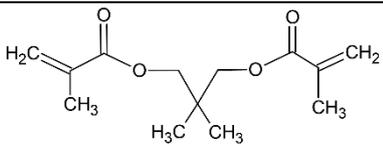 |
| Trimethylolpropane triacrylate | TMPTA | ≥ 10 | 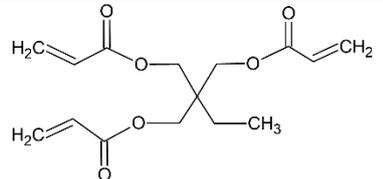 |
| Trimethylolpropane tris(3-mercaptopropionate) | TMPTP | insoluble | 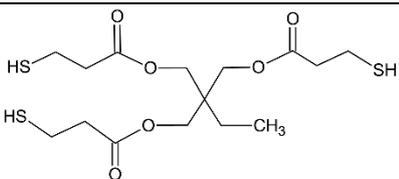 |
| Pentaerythritol tetraacrylate | PETTA | insoluble | 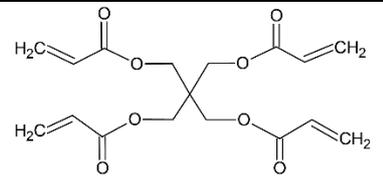 |
| Hydroxyethyl methacrylate | HEMA | ≥ 10 | 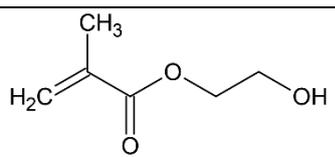 |
| Triethylene glycol dimethacrylate | TEGDMA  n = 3 | ≥ 10 | 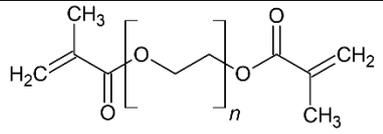 |
| Polyethylene glycol diacrylate | PEGDA250 n ~ 4 | ≥ 10 | 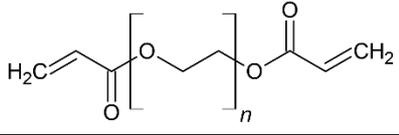 |
| Polyethylene glycol diacrylate | PEGDA700 n ~ 14 | insoluble | 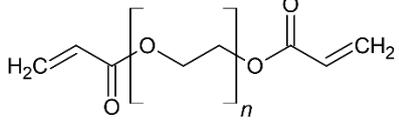 |